\documentclass[a4paper, 11pt]{article}
\usepackage{comment} 
\usepackage{booktabs}
\usepackage[nointegrals]{wasysym}
\usepackage{fullpage} 
\usepackage{graphicx}
\usepackage{cite}
\usepackage{multicol}
\usepackage[perpage]{footmisc} 
\usepackage{sectsty}
\sectionfont{\large}
\usepackage{yhmath}
\usepackage{bm}
\usepackage{float}
\usepackage{amsmath}
\usepackage[utf8]{inputenc}
\usepackage{amssymb}
\usepackage{mathrsfs}
\usepackage{wrapfig}
\usepackage{enumitem}
\usepackage[english]{babel}
\usepackage{comment}
\usepackage{bbold}  
\usepackage[usenames,dvipsnames]{xcolor} 
\usepackage{xcolor,hyperref}
\usepackage{setspace}
\usepackage{abstract}
\usepackage{tablefootnote}

\usepackage[T1]{fontenc}
\usepackage[margin=0.94in]{geometry}
\newcommand{\ra}[1]{\renewcommand{\arraystretch}{#1}}

\usepackage{enumitem}

\begin{document}
\null\hfill 
IPMU25-0046\\ 
\null\hfill LMU-ASC 22/25 \\ 
\null\hfill MPP-2025-187 \\
\vspace*{\fill}

\vspace*{\fill}
\begin{center}
    \Large\textbf{{ \textcolor{Black}{Aspects of non-minimally coupled curvature with power laws}}}

\end{center}

\begin{center}
  \normalsize\textsc{Anamaria Hell,$^{1,2}$ Dieter Lüst,$^{3,\;4}$ }
\end{center}

\begin{center}
    $^{1}$ \textit{Kavli IPMU (WPI), UTIAS,\\ The University of Tokyo,\\ Kashiwa, Chiba 277-8583, Japan}\\
    $^{2}$ \textit{Center for Data-Driven Discovery, Kavli IPMU (WPI), UTIAS,\\ The University of Tokyo, Kashiwa, Chiba 277-8583, Japan}\\
    $^{3}$\textit{Arnold Sommerfeld Center for Theoretical Physics,\\
Ludwig–Maximilians–Universität München\\
Theresienstraße 37, 80333 Munich, Germany}\\
$^{4}$\textit{Max–Planck–Institut für Physik (Werner–Heisenberg–Institut)\\
Boltzmannstra{\ss}e 8, 85748 Garching, Germany}\\ 

\end{center}
\thispagestyle{empty} 

\renewcommand{\abstractname}{\textit{\textcolor{Black}{Abstract}}}

\begin{abstract}
We consider a class of theories containing power-law terms in both the Ricci scalar and a scalar field, including their non-minimal couplings. As a first step, we systematically classify all non-trivial cases with a propagating scalar field that arise from the simplest general power-law formulation, which contains the minimal number of terms. We then analyze each case in detail, focusing on the structure of the degrees of freedom, by both formulating the theories in the Einstein frames and focusing on the singular points in the Jordan frame.  We demonstrate that such theories can give rise to different, and sometimes unexpected structure of the modes, that can change at the leading order depending on the background.
  \end{abstract}
 
\vfill
\small
\noindent\href{mailto:anamaria.hell@ipmu.jp}{\text{anamaria.hell@ipmu.jp}}\\
\href{luest@mppmu.mpg.de}{luest@mppmu.mpg.de}\\
\vspace*{\fill}

\clearpage
\pagenumbering{arabic}
\newpage

\newpage

\tableofcontents
\newpage

\section{Introduction}

Einstein's theory of General Relativity (GR) is one of the simplest and elegant descriptions of gravity, thoroughly tested across a wide range of length scales. Nevertheless, it may not be complete. The standard $\Lambda$CDM cosmological model is increasingly challenged by the recent observations and cosmological tensions \cite{ CosmoVerseNetwork:2025alb}. Among them, particularly intriguing are the Dark Energy Spectroscopic Instrument (DESI) results \cite{DESI:2025zgx}, which suggest that dark energy evolves over time, potentially including a phantom-like equation of state. Furthermore, from the perspective of quantum field theory, GR is not renormalizable beyond the one-loop corrections \cite{tHooft:1974toh, Goroff:1985th}. These considerations motivate the exploration of alternative theories of gravity that could possibly address the current challenges and open new avenues for probing the dynamics of our Universe.

Among the possible modifications, theories in which the curvature enters through power-law terms are particularly intriguing. A prominent example is the Starobinsky model, consisting of terms linear and quadratic in the Ricci scalar, which is a well-known model of inflation -- an early accelerated stage in our Universe \cite{Starobinsky:1980te, Mukhanov:1981xt} (see also \cite{Sato:1980yn, Guth:1980zm, Linde:1981mu, Chibisov:1982nx, Albrecht:1982wi,  Linde:1983gd, Mukhanov:1985rz, Sasaki:1986hm, Linde:1986fd}, and \cite{ Kallosh:2025ijd} for the review on the recent state of the inflationary models). While it is in excellent agreement with the Planck data \cite{Planck:2018jri}, it has become seemingly less favoured in the light of the recent  Atacama Cosmology Telescope (ACT) \cite{ACT:2025fju, ACT:2025tim} and South Pole Telescope (SPT) data releases \cite{SPT-3G:2025bzu}, which indicates higher values for the spectral index when combined with the Planck results, and the baryon acoustic oscillations (BAO) data from DESI. While, as pointed out in \cite{Ferreira:2025lrd}, one should take the high-$n_s$ values with caution before the CMB -- BAO tension is resolved, the possible discrepancy has nevertheless motivated several extensions of the Starobinsky inflation, by incorporating higher-order curvature terms, such as those including cubic powers or higher of the Ricci scalar \cite{Gialamas:2025ofz, Addazi:2025qra, Cheong:2025vmz}, and non-minimal couplings with a scalar field \cite{Wang:2025dbj}. 

To date, the power-law couplings between the Ricci scalar and the scalar field, including also their powers alone, have been of central interest in cosmology, modified theories of gravity, string theory and super-gravity. Non-linear powers in curvature terms naturally arise from quantum fluctuations of matter fields \cite{Sakharov:1967pk, Visser:2002ew}, when gravity is taken into account, and are expected to arise from the extra dimensions, in models
such as string theory, supergravity, or M theory  \cite{Fradkin:1985am, Nojiri:2003rz, Alvarez-Gaume:2015rwa}. While the presence of Riemann and Ricci tensors can give rise to ghost instabilities\footnote{See also a possible way to remove them with boundary conditions on an example of Conformal Gravity \cite{Maldacena:2011mk, Anastasiou:2016jix, Anastasiou:2020mik, Hell:2023rbf}.} \cite{Stelle:1977ry}, the same is not necessarily true for the terms involving the Ricci scalar (which, in the pure scale-invariant case, evaluated in flat-background, can even have no propagating modes, as shown in \cite{Hell:2023mph, Hell:2025wha} for the leading order contributions). As a result, terms with non-linear powers in the Ricci scalar were extensively discussed as possible candidates to underpin early- and late-time acceleration of the Universe \cite{Starobinsky:1980te, Maeda:1987xf, Barrow:1988xh, Muller:1989rp, Ellis:1998gf, Carroll:2004de, Capozziello:2003tk, Capozziello:2003gx, Saidov:2006xr, Nojiri:2003wx, Nojiri:2003ft,  Gunther:2004ht,Kaneda:2010ut, Kaneda:2010qv, Saidov:2010wx, Huang:2013hsb, Sebastiani:2013eqa, Artymowski:2014gea, Asaka:2015vza }, in the context of super-gravity, especially with its relation to inflation, and possible\footnote{At the moment, as pointed out in \cite{Lust:2023zql}, seemingly unlikely for the Starobinsky inflation, within the context of the String Swampland Programme \cite{Ooguri:2006in, Palti:2019pca, vanBeest:2021lhn}, due to the species scale -- scale at which the quantum gravitational effects become important \cite{Dvali:2007hz, Dvali:2007wp, Dvali:2009ks, Dvali:2010vm, Dvali:2012uq }(see also, \cite{Lehnert:2025izp} for the most recent, and probably largest Swampland review).} compatibility with string theory, when viewed as an effective field theory \cite{Cecotti:1987sa, Farakos:2013cqa, Ferrara:2013kca, Ketov:2013dfa, Ozkan:2014cua, Lust:2023zql, Gialamas:2025ofz,  Addazi:2025qra}. Notably, higher powers of Ricci scalar, excluding the linear term, are also required if one demands scale-invariance $d\geq 4$ dimensions\textcolor{Black}{ \cite{Deser:2003up, Deser:2007vs, Kounnas:2014gda, Kehagias:2015ata, Alvarez-Gaume:2015rwa, Hell:2023mph,  Hell:2025wha}. }

Models that involve the non-minimal coupling between the Ricci scalar and the scalar field are extensively studied as well, including their supersymmetric extensions. They admit inflationary and dark-energy solutions, primordial black hole formation and induced gravitational waves \cite{Futamase:1987ua, Cervantes-Cota:1995ehs, Bezrukov:2007ep, Einhorn:2009bh, Ferrara:2010yw, Ferrara:2010in, Bezrukov:2010jz, Pallis:2011gr, Arai:2011aa,  Einhorn:2012ih,Arai:2012em, Kawai:2014doa, Kawai:2014gqa,  Kawai:2015ryj, Wang:2017fuy,Pi:2017gih, He:2018gyf,  Ghilencea:2018rqg,  Gorbunov:2018llf,  Gundhi:2018wyz, Cheong:2019vzl, Bezrukov:2019ylq, Gundhi:2020zvb, Cheong:2022gfc,Kawai:2022emp,  Hyun:2023bkf,  Wang:2024vfv, Ye:2024ywg, Kim:2025dyi, Wang:2025dbj, Haque:2025uis, Kim:2025ikw, Odintsov:2025eiv}, and contain, among many prospective models, the well-known Higgs inflation \cite{Bezrukov:2007ep} (see eg. \cite{Rubio:2018ogq, Cheong:2021vdb} for a review), and also its combination with the $R^2$ inflation, known as the Higgs -- $R^2$ model  \cite{Salvio:2015kka, Calmet:2016fsr, Ema:2017rqn}. An interesting alternative to these theories where the term linear in the Ricci scalar alone is absent, is the well-known Brans-Dicke gravity \cite{Brans:1961sx}, as well as its extensions with arbitrary function of the propagating scalar and its kinetic term known as the Variable Gravity \cite{Wetterich:1987fk, Wetterich:1987fm, Wetterich:2013jsa, Wetterich:2014gaa, Wetterich:2024ieb, Hossain:2014xha},  and the no-scale Brans-Dicke gravity \cite{Hong:2025tyi, Hong:2025cae}, involving an additional  $R^2$ term and another scalar field, both intriguing as candidates to drive accelerating Universe. 

A fundamental property of all of these models is their degrees of freedom (dof) -- their building blocks, whose number is characterized by the number of initial conditions to solve the equations of motion, divided by two \cite{llandau60:mech}. Depending on the particular model, power law gravity, involving Ricci scalar and the scalar field, may thus have a different number of propagating modes. One possible way to infer their number is to explore this question directly, in the Jordan frame, where one initially formulates such theories. Another way, similarly to the Starobinsky inflation, or, more generally, f(R) gravity, is to consider formulating these theories in the Einstein frame by introducing another scalar field in a particular way, followed by a conformal transformation \cite{Whitt:1984pd, Maeda:1987xf, Barrow:1988xh, Teyssandier:1983zz, Sotiriou:2008rp, DeFelice:2010aj}, which usually presents the number of dof in a more clear way due to the Ricci scalar appearing only linearly. At the same time, such transformations do not always hold.  As pointed out in \cite{Hell:2023mph}, on an example of pure $R^2$ gravity, and in \cite{Hell:2025wha} for the d-dimensional case, if evaluated in the Jordan frame, on an Minkowski background, the pure, scale-invariant gravity does not propagate any modes, at least at the leading order. This is not obvious by looking directly in the Einstein frame, the transformation to which becomes singular for $R=0$. As a result, the change in the dof is particularly interesting from the strong-coupling perspective -- regime, in which the perturbation theory breaks down. 

As demonstrated in \cite{Hell:2023mph}, the two tensor modes and the scalar of the full $R^2$ gravity become strongly coupled due to the non-linear terms, in the limit when the parameter multiplying the linear term tends to zero. This is similar to the Vainshtein mechanism in massive gravity \cite{Vainshtein:1972sx, Dvali:2000hr,Deffayet:2001uk, Gruzinov:2001hp} -- the longitudinal mode appearing in this theory come strongly coupled at the Vainshtein scale, and decouple form the remaining dof beyond it\footnote{One should note that the vector modes become strongly coupled as well, at a scale lower than the Vainshtein radius, as shown on an example of Mimetic Massive Gravity \cite{Chamseddine:2018gqh}}, thus recovering agreement with the predictions of GR for the precession of the perihelia of Mercury and the deflection of starlight, which initially indicated a discontinuity within the linearized theory alone \cite{vanDam:1970vg, Zakharov:1970cc}. A similar case was also found in the context of massless limit of massive Yang-Mills theory, with mass introduced to the theory by hand \cite{Glashow:1961tr, Vainshtein:1971ip, Hell:2021oea}, where it was shown that due to the non-linear terms, the longitudinal modes become strongly coupled, and decouple from the remaining degrees of freedom up to small corrections that become smaller, as one approaches higher energies. To date, the strong coupling due to the appearance of modes at the leading order by modifying theories, and possible resolution in the non-linear has been found in several theories of modified gravity and massive gauge theories \cite{Dvali:2006su, Dvali:2007kt, Chamseddine:2010ub, Alberte:2010it, Mukohyama:2010xz, deRham:2010ik, deRham:2010kj, Chamseddine:2012gh, Tasinato:2014eka, Heisenberg:2014rta, BeltranJimenez:2016rff, Chamseddine:2018gqh, Heisenberg:2020xak, Hell:2021wzm, Hell:2021oea, Hell:2022wci, Hell:2023mph, Hell:2024xbv, Hell:2025uoc, Hell:2025mym}. Notably, to the different number of dof in background with $R=0$, and $R\neq0$, one could expect that the scalar and tensor modes become strongly coupled as well in the $R\to0$ limit\footnote{A. Hell would like to thank Prof. Viatcheslav Mukhanov for pointing this out during the discussion about the work \cite{Hell:2023mph}. }. 

Given the strong motivation for studying gravity theories with power-law couplings between the Ricci scalar and a scalar field -- arising both from fundamental theoretical considerations and recent observational findings -- it is essential to undertake a systematic analysis of such models in both flat and cosmological space-time backgrounds, which was missing so far. The purpose of this work is to fill this gap. By examining the general framework, as well as special cases that include a propagating scalar field, we will analyze the structure of the theories in the Einstein frame and identify the corresponding singular points in the Jordan frame. Interestingly, we find that some of these theories admit a different number of degrees of freedom, which emerge in unexpected ways.

The paper is organized in the following way: First, in Section 2, we will introduce the general power-law gravity with scalar couplings, and analyze its properties. Then, we will perform further classifications in Sections 3, 4, and 5, with focus on the non-minimal coupling between the scalar field and the Ricci tensor. In Section 6, we will consider the case when the scalar field is minimally coupled to the curvature, and finally summarize our results and conclude in Section 7.

\section{The power-law gravity with scalar couplings}
The simplest case of a general power-law gravity theory, which includes couplings between the Ricci scalar and a propagating scalar field, is given by the following action:
\begin{equation}\label{action1}
     S_{PGS}=\int d^4x\sqrt{-g}\left[\frac{M^2}{2}\left(R-2\Lambda\right)-\frac{1}{2}\nabla_{\mu}\sigma\nabla^{\mu}\sigma- \lambda \sigma^s-\frac{\xi}{2}\sigma^nR^m+\beta R^l\right].
\end{equation}
In this work, our goal is to analyze the main aspects of this theory, with focus on the degrees of freedom -- its building blocks, in flat and cosmological space-time.  As we will soon see, while this question is seemingly simple, it brings up unexpected results, particularly concerning the special cases that arise from the above action. 

One should note that this is not the most general power-law construction  --  one can, in principle, add infinite sums of scalar fields, curvature terms, or their couplings, of the form:
\begin{equation}
   U_{s}(\sigma,R)= \sum_{n,m} \left(c_n\sigma^nR^m+d_n\sigma^n+e_nR^m \right)
\end{equation}
where $c_n, d_n$ and $e_n$ are arbitrary coefficients. Moreover, one may generalize further by allowing an arbitrary potential for the scalar field, an arbitrary function of the Ricci scalar, or general couplings between the scalar field and curvature. For our purposes, however, the action (\ref{action1}) will be sufficiently general, as its special cases, which we will systematically study  in this work, will already capture the essential results.

\subsection{The map of the theory}
The action (\ref{action1}) contains several important parts: the linear and non-linear term in the Ricci scalar, the kinetic term for the scalar field $\sigma$, its self-interaction, and the non-minimal coupling with the Ricci scalar. Based on these terms, we can notice several important cases that could have a structure of the theory different from the general case.  In particular, if we limit ourselves to a propagating scalar field $\sigma$, we can notice the following special cases: 

\begin{enumerate}[label=\textbf{\textcolor{YellowOrange}{\arabic*)}}]
\item {\textsc{Einstein gravity with the non-minimally coupled scalar }}\\\\
The first possibility, that might alter the structure of the general action (\ref{action1}), is setting $\beta=0$, while keeping the remaining parameters different from zero: 
\begin{equation}\label{Enm0}
     S_{Enm}=\int d^4x\sqrt{-g}\left[\frac{M^2}{2}\left(R-2\Lambda\right)-\frac{1}{2}\nabla_{\mu}\sigma\nabla^{\mu}\sigma- \lambda \sigma^s-\frac{\xi}{2}\sigma^nR^m\right]. 
\end{equation}
This key interesting component in this action is the non-minimal coupling between the scalar field and the Ricci scalar. 
\item {\textsc{Higher-power curvature with the non-minimally coupled scalar}}\\\\
Another interesting case is to set the Einstein term to zero by setting $M=0$ in (\ref{action1}), and assuming that $l\neq 0,1$:
\begin{equation}\label{Pnm0}
     S_{Hnm}=\int d^4x\sqrt{-g}\left[-\frac{1}{2}\nabla_{\mu}\sigma\nabla^{\mu}\sigma- \lambda \sigma^s-\frac{\xi}{2}\sigma^nR^m+\beta R^l\right]. 
\end{equation}
Here, we will additionally assume that the parameters $\xi$, $\lambda$, and $\beta$ are non-vanishing. For simplicity, we have also set the cosmological constant to zero. However, one may easily reintroduce it by setting $s=0$, so it won't affect the general considerations of this case. 

\item {\textsc{Variable gravity and the power-law generalization}}\\\\
If one additionally assumes that the Ricci tensor appears only through the non-minimal coupling, by setting $M=0$, and $\beta=0$ in (\ref{action1}), and assumes that $\xi\neq0$, one can obtain the following action:  
\begin{equation}\label{vg0}
     S_{vgg}=\int d^4x\sqrt{-g}\left[-\frac{1}{2}\nabla_{\mu}\sigma\nabla^{\mu}\sigma- \lambda \sigma^s-\frac{\xi}{2}\sigma^nR^m\right]. 
\end{equation}
For $m=1$,  $n=1$, and $s=4$, the above action corresponds to the Brans-Dicke theory \cite{Brans:1961sx}, and for more general values of $n$ and $s$, the special case of variable gravity \cite{Wetterich:1987fk, Wetterich:1987fm, Wetterich:2013jsa}. In addition to studying this case, we will also consider the possibility of the Ricci scalar having higher powers.

\item {\textsc{Power-law curvature with the minimal coupling to the scalar}}\\\\
The previous cases assumed that the scalar field is non-minimally coupled to the curvature. In addition, one can also have the minimal coupling, defined in the Jordan frame, by setting $\xi=0$ in (\ref{action1}): 
\begin{equation}\label{Emin0}
     S_{minPS}=\int d^4x\sqrt{-g}\left[\frac{M^2}{2}\left(R-2\Lambda\right)-\frac{1}{2}\nabla_{\mu}\sigma\nabla^{\mu}\sigma- \lambda \sigma^s+\beta R^l\right]. 
\end{equation}

\end{enumerate}

The action (\ref{action1}) together with the special cases above assumes that the scalar field $\sigma$ is always propagating. It should be noted, however, that there is in addition a case in which its kinetic term is absent: 
\begin{equation}\label{Cnm}
     S_{Cnm}=\int d^4x\sqrt{-g}\left[\frac{M^2}{2}\left(R-2\Lambda\right)- \lambda \sigma^s-\frac{\xi}{2}\sigma^nR^m+\beta R^l\right].
\end{equation}
Due to the absence of the kinetic term, the scalar field is constrained, and thus the above action will have a completely different structure than the previous theories. The constrained scalar theory with $s=2$, $n=2$, $m=2$, and $\beta=0$ was introduced recently \cite{Hell:2025lgn}, where it was shown that for the free case, the dynamics of space-time is independent of the values of the cosmological constant. Moreover, in the presence of matter, the theory can reproduce a phantom-like behavior of the equation of state, while being healthy. Another interesting case of (\ref{Cnm}) is the Brans-Dicke theory with $\omega=0$, also known as the massive dilaton gravity \cite{OHanlon:1972xqa, Wands:1993uu, Sotiriou:2008rp}, which corresponds to setting $m=1$, $n=1$, $M=0$, and $\beta=0$. This case is connected to the $f(R)$ gravity with a frame transformation \cite{Teyssandier:1983zz}.

In this work, we will analyze only the cases where the scalar field is propagating, described by the general action (\ref{action1}), and its special cases, (\ref{Enm0}) --   (\ref{Emin0}). These theories have a very distinguished structure, which, as we will show, will affect the counting of the degrees of freedom. 

\subsection{The general theory and the Einstein frame}

As a first step to analyze the properties of the theory given in (\ref{action1}), let us first study its formulation in the Einstein frame. Here, we will assume that all coefficients $M, \xi$, and $\beta$ are non-vanishing. 

In order to write (\ref{action1}) in the Einstein frame, we will introduce a scalar field in the following way, similarly to the case of $f(R)$ gravity:
\begin{equation}\label{sgen2}
    S_{PGS,E}=\int d^4x\sqrt{-g}\left[f(\chi,\sigma)+f_{,\chi}(R-\chi)-\frac{1}{2}\nabla_{\mu}\sigma\nabla^{\mu}\sigma- \lambda \sigma^s-M^2\Lambda\right], 
\end{equation}
where $_{,\chi}$ denotes a derivative with respect to the scalar $\chi$, and 
\begin{equation}
    f(\chi)=\frac{M^2}{2}\chi-\frac{\xi}{2}\sigma^n\chi^m+\beta\chi^l.
\end{equation}
By varying with respect to $\chi$, we find the following constraint:
\begin{equation}
    f_{,\chi\chi}(\chi,\sigma)(R-\chi)=0. 
\end{equation}
This leads us to two possible solutions: either we have 
\begin{equation}
    f_{,\chi\chi}\neq0,
\end{equation}
which implies that 
\begin{equation}
    R=\chi,
\end{equation} 
or we have:  
\begin{equation}
    f_{,\chi\chi}=-\frac{\xi m (m-1)}{2}\sigma^n\chi^{m-1} + \beta l(l-1)\chi^{l-2}=0.
\end{equation}
The latter case is fulfilled by having 
\begin{equation}
   \left( m=0\quad \text{or}\quad m=1 \right), \quad \text{and} \quad\left( l=0\quad \text{or} \quad l=1 \right),
\end{equation}
taking values at the same time. Let us assume that this is not the case. Then, by substituting $\chi=R$ in (\ref{sgen2}), we recover the initial action (\ref{action1}). Furthermore, by defining:
\begin{equation}\label{conformaltransfGen}
    F= f_{,\chi},\qquad\text{and}\qquad g_{\mu\nu}=\frac{M^2}{F}\Tilde{g}_{\mu\nu},
\end{equation}
and expressing the old metric in terms of the new one, the previous action becomes: 
\begin{equation}
    \begin{split}
        S_{genE}&=\int d^4x\sqrt{-\Tilde{g}}\left\{M^2\Tilde{R}+\frac{M^4}{F^2}\left[-\frac{3}{2M^2}\Tilde{\nabla}_{\mu}F\Tilde{\nabla}^{\mu}F-\frac{ F}{2M^2}\Tilde{\nabla}_{\mu}\sigma\Tilde{\nabla}^{\mu}\sigma -M^2\Lambda-\lambda\sigma^s\right.\right.\\&\left.\left.+\frac{\xi}{2}\sigma^n\chi^m(m-1)+\beta\chi^l(1-l)\right]\right\}
    \end{split}
\end{equation}
Thus, we arrive at an action of Einstein's gravity that describes two tensor modes and the two coupled scalars, with
\begin{equation}\label{FrelationGen}
    F=\frac{M^2}{2}-\frac{\xi m}{2}\sigma^n\chi^{m-1}+\beta l \chi^{l-1}. 
\end{equation}
Even though the fields $\chi$ and $\sigma$ are coupled, it is possible to see that the theory describes two scalar modes by considering (\ref{FrelationGen}) perturbatively, as a function of $\chi$ and $\psi$, with their background values vanishing. Then, at the leading order, is we assume that $m>1$ and $l>1$, the kinetic terms of $\chi$ and $\sigma$ decouple:
\begin{equation}
    \frac{M^4}{F^2}\left[-\frac{3}{2M^2}\Tilde{\nabla}_{\mu}F\Tilde{\nabla}^{\mu}F-\frac{ F}{2M^2}\Tilde{\nabla}_{\mu}\sigma\Tilde{\nabla}^{\mu}\sigma \right]\sim -\frac{6\beta^2 l^2}{M^2}\Tilde{\nabla}_{\mu}\chi^{l-1}\Tilde{\nabla}^{\mu}\chi^{l-1}- \Tilde{\nabla}_{\mu}\sigma\Tilde{\nabla}^{\mu}\sigma
\end{equation}

We can also further simplify this action by defining:
\begin{equation}\label{OmegaDef}
    \Omega=M \ln\left(\frac{2F}{M^2}\right)
\end{equation}
after which it becomes:
\begin{equation}
     \begin{split}
           S_{genE}&=\int d^4x\sqrt{-\tilde{g}}\left\{M^2\tilde{R}-\frac{3}{2}\tilde{\nabla}_{\mu}\Omega\tilde{\nabla}^{\mu}\Omega-e^{-\Omega/M}\tilde{\nabla}_{\mu}\sigma\tilde{\nabla}^{\mu}\sigma\right.\\&\left. +4e^{-2\Omega/M}\left[-M^2\Lambda-\lambda\sigma^s+\frac{\xi}{2}\sigma^n\chi^m(m-1)+\beta\chi^l(1-l)\right]\right\},
     \end{split}
\end{equation}
where we should express $\chi$ as a function of $\Omega$ and $\sigma$. This is easy to achieve if $m=1$ or $m=0$, and $l\neq 0,1$, for which we find: 
\begin{equation}
    \chi=\left\{\frac{1}{\beta h}\left[\frac{M^2}{2}\left(e^{\Omega/M}-1\right)+\frac{\xi}{2}\sigma^n\right]\right\}^{\frac{1}{l-1}},
\end{equation}
or, if $m\neq 0, 1$, and $l=1$ or $l=0$:
\begin{equation}
    \chi=\left\{\frac{2}{\xi m \sigma^n}\left[\left(1-e^{\Omega/M}\right)\frac{M^2}{2}\right]\right\}^{\frac{1}{m-1}}
\end{equation}
 Otherwise, to use this formulation in terms of $\Omega$, one needs to solve a polynomial equation for $\chi$, from the definition of $F$, so in that case it might be even better to work in terms of $\chi$ and $\sigma$ directly.

\subsection{Singular points and the Jordan frame }
In the general case, we have seen that the power-law action (\ref{action1}) describes four degrees of freedom: two tensor modes that correspond to the gravitational waves, and two scalar modes that arise due to the non-minimal coupling with the curvature and its powers. The key to having a clear insight into this was the Einstein frame, in which the general powers of curvature were reduced to match Einstein's relativity, together with two non-trivial scalars. 

Nevertheless, the previous analysis should be taken with caution -- it does not hold everywhere. As we have seen, the conformal transformation (\ref{conformaltransfGen}) becomes singular by taking values $m=0,1$ and $l=0,1$ at the same time, and thus does not hold for these choices. However, with this choice, the possibly singular terms match with the other ones, already contained in (\ref{action1}), so thus, this analysis won't give rise to any new insights. 

Another important point is the case when $R=0$. While the conformal transformation (\ref{conformaltransfGen}) itself is not singular, it corresponds to the vanishing scalar $\chi$, and should therefore be considered with care. To demonstrate this, let us go back to the Jordan frame, with the general action (\ref{action1}), set $\Lambda=0$ and assume that $l>1$. Then, one can easily see that the Minkowski solution is trivially satisfied by the equations of motion and can be used as a background. 

To analyze the degrees of freedom in this case, we will consider small perturbations around a background:
\begin{equation}
    g_{\mu\nu}=\eta_{\mu\nu}+h_{\mu\nu},
\end{equation}
with $\eta_{\mu\nu}$ being the Minkowski metric. In this case, we find several important terms arising from the action (\ref{action1}), with the corresponding Lagrangian densities given by:
\begin{equation}
    \mathcal{L}=\mathcal{L}_{EH}+\mathcal{L}_{\sigma s} + \mathcal{L}_{n m} + \mathcal{L}_{l}, 
\end{equation}
where
\begin{equation}
    \begin{split}
        \mathcal{L}_{EH}&=\frac{M^2}{2}\sqrt{-g}R=\frac{M^2}{8}\left(-h_{\mu\nu,\alpha}h^{\mu\nu,\alpha}+2h_{\mu\nu}^{,\mu}h^{\alpha}_{\nu,\alpha}-2h^{\mu\nu}_{,\mu}h_{,\nu}+2h_{,\nu}h^{,\nu}\right)+\mathcal{O}\left(h^{3}\right)=\\
        \mathcal{L}_{\sigma s} &= \sqrt{-g}\left(-\frac{\alpha}{2}\nabla_{\mu}\sigma\nabla^{\mu}\sigma- \lambda \sigma^s\right)=-\frac{\alpha}{2}\sigma_{,\mu}\sigma^{,\mu}- \lambda \sigma^s+\mathcal{O}\left(\sigma^2h, \sigma^s h\right)\\
        \mathcal{L}_{n m} &= -\frac{\xi}{2}\sqrt{-g}\sigma^n R^m=-\frac{\xi}{2}\sigma^n (h_{\mu\nu}^{,\mu\nu}-\Box h)^m+\mathcal{O}\left(\sigma^n h^{m+1}\right)\\
        \mathcal{L}_l &= \beta\sqrt{-g} R^l=\beta(h_{,\mu\nu}^{\mu\nu}-\Box h)^l+\mathcal{O}\left(h^{l+1}\right)
    \end{split}
\end{equation}
where $h=h^{\mu}_{\mu}$ and $_{,\mu}=\partial_{\mu}$. We are primarily interested in the degrees of freedom of the linearized theory, whose action is quadratic in perturbations. While the first two Lagrangian densities contribute to this order, we can notice that $\mathcal{L}_{nm}$, and $\mathcal{L}_l$ do not necessarily have to, depending on the powers of the fields. In particular, if $l>2$, the contribution of $\mathcal{L}_l$ will not contribute this count of the degrees of freedom. Similar holds for $\mathcal{L}_{nm}$, which contributes to the count and behavior of the linearized dof only for $n=1$ and $m=1$. This thus gives rise to the difference with the Einstein frame, which we will explore in the following section.

To fully analyze the non-trivial case of the linearized theory, we will decompose the metric perturbations according to the irreducible representations of the rotation group:
\begin{equation}\label{decompositionFlat}
    \begin{split}
        h_{00}&=2\phi\\
        h_{0i}&=S_i+B_{,i}\\
        h_{ij}&=2\psi\delta_{ij}+E_{,ij}+F_{i,j}+F_{j,i}+h_{ij}^T,
    \end{split}
\end{equation}
where
\begin{equation}
    S_{i,i}=0,\qquad F_{i,i}=0,\qquad h_{ii}=0\qquad \text{and}\qquad h^T_{ij,i}=0. 
\end{equation}
At the leading order, the scalar, vector, and tensor perturbations decouple. The simplest cases are the vector and tensor modes, which arise only from $\mathcal{L}_{EH}$ at the leading order in perturbations. For the tensor modes, we find:
\begin{equation}
   \mathcal{L}_T=-\frac{M^2}{8}h_{ij,\mu}^Th^{,\mu}_{ij}.
\end{equation}
For the vector modes, the Lagrangian density is given by:
\begin{equation}
    \mathcal{L}_V=-\frac{M^2}{4}V_i\Delta V_i,
\end{equation}
where $V_i=S_i-\dot{F}_i$ is the gauge invariant combination. By varying the action with respect to $V_i$, we find:
\begin{equation}
    \Delta V_i=0,
\end{equation}
and thus, by substituting this constraint back to the Lagrangian, we find:
\begin{equation}
    \mathcal{L}_V=0,
\end{equation}
meaning that the vector modes do not propagate. 
For the scalar modes, we find a slightly more complicated Lagrangian density:
\begin{equation}\label{langGenSc}
    \begin{split}
        \mathcal{L}_S&=M^2\left[2\phi\Delta\psi-3\dot{\psi}^2-\psi\Delta\psi\right]-\frac{\alpha}{2}\sigma_{,\mu}\sigma^{,\mu}-\frac{\xi}{2}2^m\sigma^n(\Delta\phi+3\Ddot{\psi}-2\Delta\psi)^m-\lambda \sigma^s\\&+2^l\beta\left(\Delta\phi+3\Ddot{\psi}-2\Delta\psi\right)^l
    \end{split}
\end{equation}
Depending on the values of $n,m,s$ and $l$, we will have different contributions, and thus a different number of modes. Let's analyze them on a case-by-case basis. Similarly, we can set $s=2$ to keep the contribution arising from only the scalar mode $\sigma$. 

\begin{center}
    \textit{\textbf{Case 1:} $m=1$, $n=1$ and $l=2$}
\end{center}

If $m=n=1$, and $l=2$, both non-minimal coupling and higher powers in curvature contribute to the scalar sector. Thus, (\ref{langGenSc}) becomes: 
\begin{equation}\label{langGenSc1}
    \begin{split}
        \mathcal{L}_S=M^2\left[2\phi\Delta\psi-3\dot{\psi}^2-\psi\Delta\psi\right]-\frac{1}{2}\sigma_{,\mu}\sigma^{,\mu}-\xi\sigma(\Delta\phi+3\Ddot{\psi}-2\Delta\psi)-\lambda \sigma^s+4\beta\left(\Delta\phi+3\Ddot{\psi}-2\Delta\psi\right)^2
    \end{split}
\end{equation}
We can notice that $\phi$ is not propagating -- it does not show in a form $\dot{\phi}^2$, in contrast to the other scalars. By varying the action with respect to it, we find the following constraint: 
\begin{equation}
    8\beta\Delta\phi=-\left[8\beta(3\Ddot{\psi}-2\Delta\psi)-\xi\sigma+2M^2\psi\right]
\end{equation}
By solving it, and substituting back into (\ref{langGenSc1}), we find: 
\begin{equation}
    \mathcal{L}_S=-\frac{1}{2}\sigma_{,\mu}\sigma^{,\mu}-\frac{\xi^2}{16\beta}\sigma^2-\lambda \sigma^2+M^2\left(3\dot{\psi}^2+3\psi\Delta\psi-\frac{M^4}{4\beta}\psi^2\right)+\frac{\xi M^2}{4\beta}\sigma\psi.
\end{equation}
Therefore, in this case, the theory describes two scalar modes $\sigma$ and $\psi$. Interestingly, we can notice that both scalars have acquired a mass which is inversely proportional to $\beta$, and are linearly coupled.

\begin{center}
    \textit{\textbf{Case 2:} $m>1$, or  $n>1$, and $l=2$}
\end{center}

If either $m$, or $n$ becomes larger than unity, then the non-minimal coupling no longer contributes at the linearized order, and we find:
\begin{equation}
    \begin{split}
        \mathcal{L}_S=M^2\left[2\phi\Delta\psi-3\dot{\psi}^2-\psi\Delta\psi\right]-\frac{1}{2}\sigma_{,\mu}\sigma^{,\mu}-\lambda \sigma^2+4\beta\left(\Delta\phi+3\Ddot{\psi}-2\Delta\psi\right)^2
    \end{split}
\end{equation}
The scalar $\phi$ still does not propagate, and satisfies the following constraint: 
\begin{equation}
    8\beta\Delta\phi=-\left[8\beta(3\Ddot{\psi}-2\Delta\psi)+2M^2\psi\right]. 
\end{equation}
After solving it, and substituting back to the action, we find: 
\begin{equation}
    \mathcal{L}_S=-\frac{1}{2}\sigma_{,\mu}\sigma^{,\mu}-\lambda \sigma^2+M^2\left(3\dot{\psi}^2+3\psi\Delta\psi-\frac{M^4}{4\beta}\psi^2\right)
\end{equation}
Thus, similarly to the previous case, we again find two propagating scalars, which are now decoupled. In addition, we can notice that the mass of the scalar $\sigma$ is now not modified, as it was the case with the non-minimal coupling. 

\begin{center}
    \textit{\textbf{Case 3:} $m=1$,   $n=1$, and $l>2$}
\end{center}
For $l>2$, the $R^l$ term no longer contributes ot the Lagrangian density at the linearized order. The above choice for $n$ and $m$, in contrast, still secures the contribution that arises from the non-minimal coupling. Thus, in this case, the Lagrangian density describing the scalar modes is given by: 
\begin{equation}
    \begin{split}
        \mathcal{L}_S=M^2\left[2\phi\Delta\psi-3\dot{\psi}^2-\psi\Delta\psi\right]-\frac{1}{2}\sigma_{,\mu}\sigma^{,\mu}-\xi\sigma(\Delta\phi+3\Ddot{\psi}-2\Delta\psi)-\lambda \sigma^2
    \end{split}
\end{equation}
By varying with respect to $\phi$, we now find a constraint, whose solution is given by:
\begin{equation}
    \psi=\frac{\xi}{2M^2}\sigma.
\end{equation}
By substituting this back to the action, we then obtain a single scalar mode: 
\begin{equation}
    \mathcal{L}_S=-\left(\frac{1}{2}+\frac{3\xi^2}{4 M^2}\right)\sigma_{,\mu}\sigma^{,\mu}-\lambda\sigma^2
\end{equation}

\begin{center}
    \textit{\textbf{Case 4:} $m>1$, or  $n>1$, and $l>2$}
\end{center}

Finally, in the last case, the only contribution comes from the Einstein term, which is linear in the Ricci scalar:
\begin{equation}
    \begin{split}
        \mathcal{L}_S=M^2\left[2\phi\Delta\psi-3\dot{\psi}^2-\psi\Delta\psi\right]-\frac{1}{2}\sigma_{,\mu}\sigma^{,\mu}-\lambda\sigma^2
    \end{split}
\end{equation}
By varying with respect to the $\phi$, we find:
\begin{equation}
    \Delta\psi=0,
\end{equation}
which implies: 
\begin{equation}
    \psi=0.
\end{equation}
Thus, only one scalar mode remains in the theory:
\begin{equation}
    \begin{split}
        \mathcal{L}_S=-\frac{1}{2}\sigma_{,\mu}\sigma^{,\mu}-\lambda\sigma^2,
    \end{split}
\end{equation}
which was also initially assumed.

To summarize, we have found that the general theory for the $R\neq 0$ case describes two scalar and two tensor modes. However, the $R=0$ point is singular, due to the vanishing of the scalar $\chi$, and thus, one cannot use the Einstein frame in this case. In particular, by studying the Minkowski case, we have found that unless $n=1$, $m=1$, and $l=2$, this theory for $R=0$ in the Jordan frame describes one massive scalar mode. Therefore, in this case, the Einstein frame is not good enough to make conclusions, despite the conformal transformation being regular.

\section{Einstein gravity with the non-minimally coupled scalar}
In the previous section, we have seen that in general, the theory of gravity with powers in the Ricci scalar and non-minimally coupled scalar field propagates two scalar and two tensor modes. This also persists in the Jordan frame around flat space-time (with $\Lambda=0$), where it also contains two coupled scalar modes, if $l=2$ and $n=m=1$. However, the situation changes for other cases of $l,\;n$, and $m$. 

In this section, continue this exploration on a special case of the power-law gravity with scalars, focusing now on the non-minimal term, and setting the non-linear curvature contributions to zero with $\beta=0$. In this case, the theory is described by Einstein gravity, with a non-minimally coupled scalar field: 
\begin{equation}\label{Enm}
     S_{Enm}=\int d^4x\sqrt{-g}\left[\frac{M^2}{2}\left(R-2\Lambda\right)-\frac{1}{2}\nabla_{\mu}\sigma\nabla^{\mu}\sigma- \lambda \sigma^s-\frac{\xi}{2}\sigma^nR^m\right],  
\end{equation}
in which we will assume that $M\neq0$,  and $\xi\neq 0$.Similarly to the previous case, let us first analyze the theory in general, with the help of the Einstein-frame formulation, and then focus on the points at which this transformation appeared to be singular. 

\subsection{The general analysis}
As a first step, let's introduce a scalar field $\chi$: 
\begin{equation}\label{mnccurv2}
    S_{Enm}=\int d^4x\sqrt{-g}\left[f(\chi,\sigma)+f_{,\chi}(R-\chi)-\frac{1}{2}\nabla_{\mu}\sigma\nabla^{\mu}\sigma-M^2\Lambda- \lambda \sigma^s\right], 
\end{equation}
where
\begin{equation}
    f(\chi)=\frac{M^2}{2}\chi-\frac{\xi}{2}\sigma^n\chi^m.
\end{equation}
By varying the action with respect to it, we find the following constraint: 
\begin{equation}
   \xi\sigma^n m (m-1)\chi^{m-2}(R-\chi)=0. 
\end{equation}
We can see that 
\begin{equation}
    \chi=R 
\end{equation}
holds only if
\begin{equation}
    \chi\neq 0,\qquad \sigma\neq 0\qquad\text{or}\qquad m=0,1.
\end{equation}
While the case with $m=0$ is trivial, the remaining possibilities should be taken with caution, and we will return to them in the following subsection. For now, let's assume that $f_{,\chi\chi}\neq0$. Then, by performing the conformal transformation:
\begin{equation}
    g_{\mu\nu}=\frac{M^2}{F}\Tilde{g}_{\mu\nu}, \qquad F=\frac{M^2}{2}-\frac{\xi}{2}m\sigma^n\chi^{m-1}, 
\end{equation}
we find:
\begin{equation}\label{mnccurvE}
    \begin{split}
        S_{Enm}&=\int d^4x\sqrt{-\Tilde{g}}\left\{M^2\Tilde{R}+\frac{M^4}{F^2}\left[-\frac{3}{2M^2}\Tilde{\nabla}_{\mu}F\Tilde{\nabla}^{\mu}F-\frac{F}{2M^2}\Tilde{\nabla}_{\mu}\sigma\Tilde{\nabla}^{\mu}\sigma\right.\right.\\&\left.\left. -M^2\Lambda- \lambda \sigma^s+\frac{\xi}{2}\sigma^n\chi^m(m-1)\right]\right\}
    \end{split}
\end{equation}
By further defining:
\begin{equation}
    \Omega=M\ln\left(\frac{2F}{M^2}\right)
\end{equation}
we find:
\begin{equation}
  \begin{split}
       S_{Enm}&=\int d^4x\sqrt{-\Tilde{g}}\left\{M^2\Tilde{R}-\frac{3}{2}\Tilde{\nabla}_{\mu}\Omega\Tilde{\nabla}^{\mu}\Omega-  e^{-\frac{\Omega}{M}}\Tilde{\nabla}_{\mu}\sigma\Tilde{\nabla}^{\mu}\sigma\right.\\&\left.+4e^{-\frac{2\Omega}{M}}\left[-M^2\Lambda- \lambda \sigma^s+\frac{m-1}{2}\sigma^{\frac{n}{1-m}}\left(\frac{M^2}{\xi m}\left(1-e^{\frac{\Omega}{M}}\right)\right)^{\frac{1}{m-1}}\right]\right\}.
  \end{split}
\end{equation}
In the general case, the above action clearly describes two scalar modes -- $\Omega$ and $\sigma$ -- and two tensor modes that arise from $\tilde{R}$. 
 
Let us now focus on the special cases for which the previous analysis does not hold. We have previously seen that this happens if $m=1$, or if $\chi$ or $\sigma$ are zero. For this, we will thus go back to the Jordan frame. 

\subsection{The Jordan frame}
Now, we will focus on the singular points of the theory (\ref{Enm}), starting first, with flat space-time considerations, followed by the $m=1$ analysis. 
\subsubsection{The flat space-time}
Let us first go back to the original action, and for simplicity set $\Lambda=0$. In this case, the flat metric is a solution to the equations of motion. Following the analysis of the general theory, let's perturb around the flat space-time, and decompose the metric according to (\ref{decompositionFlat}). Due to the Einstein term, we will find two tensor modes, with the Lagrangian density given by:
\begin{equation}
   \mathcal{L}_T=-\frac{M^2}{8}h_{ij,\mu}^Th^{,\mu}_{ij}, 
\end{equation}
and no propagating vector modes. Similarly to the general theory, the scalar modes will again depend on the powers of the curvature, and the scalar: 
\begin{equation}
    \mathcal{L}_S=M^2\left(2\phi\Delta\psi-3\dot{\psi}^2-\psi\Delta\psi\right)-\frac{1 }{2}\partial_{\mu}\sigma\partial^{\mu}\sigma-\frac{\xi}{2}2^m\sigma^n\left(\Delta\phi+3\ddot{\psi}-2\Delta\psi\right)^{m}-\lambda\sigma^s,
\end{equation}
thus leading us to several different cases. To have a non-trivial contribution at leading order, we will set $s=2$, which, as we will see, cannot affect the number of dof for this theory. 

\begin{equation}
    \textit{\textbf{Case 1:} $n>1$ or $m>1$}    
\end{equation}
In this case, the leading order contributions to the Lagrangian density are given by: 
\begin{equation}
    \mathcal{L}_S=M^2\left(2\phi\Delta\psi-3\dot{\psi}^2-\psi\Delta\psi\right)-\frac{1 }{2}\partial_{\mu}\sigma\partial^{\mu}\sigma-\lambda\sigma^2.
\end{equation}
We can notice that $\phi$ does not propagate, and gives rise to the following constraint:
\begin{equation}
    \Delta\psi=0, 
\end{equation}
whose solution is given by:
\begin{equation}
    \psi=0.
\end{equation}
Thus, in this case, the theory will describe a single scalar mode: 
\begin{equation}
    \mathcal{L}_S=-\frac{1 }{2}\partial_{\mu}\sigma\partial^{\mu}\sigma-\lambda\sigma^2. 
\end{equation}

\begin{equation}
    \textit{\textbf{Case 2:} $n=1$ and $m=1$}    
\end{equation}

In this case, the constraint for $\phi$ changes, giving rise to the following relation:
\begin{equation}
    \xi\sigma=2M^2\psi. 
\end{equation}
By expressing $\sigma$ in terms of the potential $\psi$, we then find:
\begin{equation}
  \mathcal{L}_S=-\frac{3M^2\xi^2+2  M^4}{\xi^2}\partial_{\mu}\psi\partial^{\mu}\psi
\end{equation}
Therefore, we arrive at a single scalar mode, which is not a ghost if the following condition is fulfilled: 
\begin{equation}
    3\xi^2+2  M^2>0. 
\end{equation}
\subsubsection{Homogeneous and isotropic Universe with $m=1$}
Previously, we have seen that the flat space-time changes the number of degrees of freedom in the Jordan frame. This analysis was performed both for $m>1$ and $m=1$. Therefore, the last singular point that we have to evaluate is the case with $R\neq 0$, but $m=1$. 

Let us consider this theory in a cosmological background: 
\begin{equation}\label{FLRWbcg}
    ds^2=-N(t)^2dt^2+a(t)^2\delta_{ij}dx^idx^j,
\end{equation}
with $N$ being the lapse, and assume that $n\neq1$. To study the dof, we will next perturb around (\ref{FLRWbcg}): 
\begin{equation}
    g_{\mu\nu}=g_{\mu\nu}^{(0)}+\delta g_{\mu\nu},
\end{equation}
and decompose the perturbations according to the irreducible representations of the rotation group: 
\begin{equation}
    \begin{split}
        \delta g_{00}&=-2\phi\\
        \delta g_{0i}&=a^2(t)\left(S_i+B_{,i}\right)\\
        \delta g_{ij}&=a^2(t)\left(2\psi\delta_{ij}+2 E_{,ij}+F_{i,j}+F_{j,i}+h_{ij}^T\right). 
\end{split}
\end{equation}
In the above relations, and similarly to the flat space-time, the vector and tensor modes satisfy:
\begin{equation}
    S_{i,i}=0,\qquad F_{i,i}=0,\qquad h_{ij,i}^T=0,\qquad\text{and}\qquad h_{ii}^T=0. 
\end{equation}
In addition to the background metric, we will assume that the scalar field has a non-vanishing background value as well, and perturb around it:
\begin{equation}\label{FLRWsc}
    \phi=\sigma(t)+\delta\sigma. 
\end{equation}
Without the loss of generality, let us set $\lambda=0$. 
By varying the action with respect to the lapse, we find: 
\begin{equation}
    6 \xi  n \sigma^{n -1} \dot{\sigma} a \dot{a}+6 \dot{a}^{2} \sigma^{n} \xi +\left(   \dot{\sigma}^{2}+2 \Lambda \,M^{2}\right) a^{2}-6 M^{2} \dot{a}^{2}
=0
\end{equation}
Next, by varying with respect to the scale factor, we find:
\begin{equation}
    \begin{split}
        &a^{2} \xi  n \left(n -1\right) \sigma^{n -2} \dot{\sigma}^{2}+2 \left(\dot{a} \dot{\sigma}+\frac{\ddot{\sigma} a}{2}\right) n a \xi  \sigma^{n -1}+\xi  \left(\dot{a}^{2}+2 \ddot{a} a\right) \sigma^{n}\\&+\frac{\left(-   \dot{\sigma}^{2}+2 \Lambda \,M^{2}\right) a^{2}}{2}-2 \ddot{a} M^{2} a-M^{2} \dot{a}^{2}
=0
    \end{split}
\end{equation}

Finally, by varying with respect to the scalar field, we find:
\begin{equation}
    3\xi  \left(\ddot{a} a+\dot{a}^{2}\right) n \sigma^{n -1}+   a \left(\ddot{\sigma} a+3 \dot{a} \dot{\sigma}\right)=0
\end{equation}

The above three equations can be solved for $\{\Lambda,\; \ddot{a},\; \ddot{\sigma}\}$, and we will assume that they are always satisfied when analyzing the perturbations. 

To find the degrees of freedom, we expand the action up to second order in perturbations. At this order, the scalar, vector, and tensor dof decouple. 

Let us consider the scalar modes. We will work in the longitudinal gauge, in which $E=0$ and $B=0$, while the remaining scalars coincide with the gauge-invariant variables \cite{Bardeen:1980kt, Mukhanov:1990me, Mukhanov:2005sc}. After integrating the action several times by parts, we find that the scalar $\phi$ is not propagating. We find its constraint, solve it, and substitute it back to the action. After this, the action becomes a function of just two fields -- $\psi$ and $\delta\sigma$. However, the determinant of the kinetic matrix associated with the two modes vanishes, meaning that the two are not independent, even though both come with terms of the form $\dot{\psi}^2$, and $\dot{\delta\sigma}^2$, together with a mixed term as well. We then perform a substitution:
\begin{equation}
    \delta\sigma=\delta\sigma_2+\psi \frac{a\dot{\sigma}}{\dot{a}},
\end{equation}
after which, the $\dot{\psi}^2$ term cancels from the action, and allows us to find the constraint for psi, solve it, and substitute back into the action. Thus, we remain with a single scalar mode $\delta\psi_2$, whose no-ghost condition is given by:
\begin{equation}
\text{NG}=\frac{3 \dot{a}^{2} \left(\xi^{2} n^{2} \sigma^{2 n -2}+\frac{2    \left(M^{2}-\xi  \sigma^{n}\right)}{3}\right) \sigma^{2} a^{3} \left(M^{2}-\xi  \sigma^{n}\right)}{\left(a n \sigma^{n} \dot{\sigma} \xi -2 M^{2} \dot{a} \sigma+2 \xi  \sigma^{n +1} \dot{a}\right)^{2}}
\end{equation}

and the speed of propagation in the limit of high energies is unity:
\begin{equation}
    c_S^2=1. 
\end{equation}

The previous analysis can also be performed in the Einstein frame (with $n\neq 1$). By defining a metric:
\begin{equation}
    g_{\mu\nu}=\frac{M^2}{\Sigma}g\Tilde{g}_{\mu\nu},\qquad \text{where}\qquad \Sigma=\frac{M^2}{2}-\frac{\xi}{2}\sigma^n, 
\end{equation}
we find a single scalar degree of freedom:
\begin{equation}
    S=\int d^4x \sqrt{-\Tilde{g}}\left[M^2 \Tilde{R}-\frac{1}{2}\Tilde{\nabla}_{\mu}\sigma\Tilde{\nabla}^{\mu}\sigma\left(\frac{3M^2\xi^2 n^2}{4\Sigma^2}\sigma^{2n-2}\right)-\frac{ \Sigma}{M^2}\right]
\end{equation}
which further confirms the analysis in the Jordan frame. Moreover, it also makes the tensor modes manifest, which are also clearly contained in the Jordan frame, due to the terms linear in the Ricci scalar.

\section{Higher-power curvature with the non-minimally coupled scalar}
As a next case, let's set $M=0$, having a generalization of pure gravity, with non-minimal coupling:  
\begin{equation}\label{Pnm}
     S_{Hnm}=\int d^4x\sqrt{-g}\left[-\frac{1 }{2}\nabla_{\mu}\sigma\nabla^{\mu}\sigma- \lambda \sigma^s-\frac{\xi}{2}\sigma^nR^m+\beta R^l\right], 
\end{equation}
with $l\neq 0, 1$. Here,  we have set $\Lambda=0$ without the loss of generality. Similarly to the previous cases, for $R\neq 0$, this theory can be written in the Einstein frame: 
\begin{equation}
    \begin{split}
        S_{Hnm,E}&=\int d^4x\sqrt{-\Tilde{g}}\left\{M^2\Tilde{R}+\frac{M^4}{F^2}\left[-\frac{3}{2M^2}\Tilde{\nabla}_{\mu}F\Tilde{\nabla}^{\mu}F-\frac{  F}{2M^2}\Tilde{\nabla}_{\mu}\sigma\Tilde{\nabla}^{\mu}\sigma -\lambda\sigma^s\right.\right.\\&\left.\left.+\frac{\xi}{2}\sigma^n\chi^m(m-1)+\beta\chi^l(l-1)\right]\right\},
    \end{split}
\end{equation}
where
\begin{equation}
    F=-\frac{\xi m}{2}\sigma^n\chi^{m-1}+\beta l \chi^{l-1}. 
\end{equation}
This result can be found by performing the analysis analogous to the general case. Similarly to it, we can see that this theory propagates again propagates four degrees of freedom: the two scalar modes, and the two tensor ones.

Let us now consider the case $R=0$, assuming $\lambda=0$, without the loss of generality. In this case, the previous analysis does not hold, as it implies that $\chi=0$. As a special case, we will again consider the flat space-time. To study the perturbations, we decompose the metric according to (\ref{decompositionFlat}). Now, at the leading order, we can clearly see that there are no tensor modes at leading order, similarly to the pure $R^2$ gravity. They usually arise due to the linear term in the flat space-time. Moreover, one can easily see that the vector modes are not propagating as well. Thus, the only non-trivial part corresponds to the scalar modes, whose Lagrangian density at leading order is given by: 
\begin{equation}\label{Lspure}
    \mathcal{L}_S=-\frac{ 1}{2}\partial_{\mu}\sigma\partial^{\mu}\sigma-\frac{\xi}{2}2^m\sigma^n\left(\Delta\phi+3\ddot{\psi}-2\Delta\psi\right)^{m}+2^l\beta\left(\Delta\phi+3\Ddot{\psi}-2\Delta\psi\right)^l
\end{equation}
Similarly to the previous sections, we will have different cases depending on the values of $n, m$, and $l$.

\begin{equation}
    \textit{\textbf{Case 1:} $n=1$, $m=1$ and  $l=2$}    
\end{equation}
In this case, both non-minimal coupling and the $R^2$ term contribute to the scalar modes. Thus, (\ref{Lspure}) becomes:
\begin{equation}
    \mathcal{L}_S=-\frac{ 1}{2}\partial_{\mu}\sigma\partial^{\mu}\sigma-\xi\sigma\left(\Delta\phi+3\ddot{\psi}-2\Delta\psi\right)+4\beta\left(\Delta\phi+3\Ddot{\psi}-2\Delta\psi\right)^2.
\end{equation}
We can notice that the scalar $\phi$ is again constrained, satisfying the following relation: 
\begin{equation}
    8\beta\Delta\phi=-[8\beta(3\ddot{\psi}-2\Delta\psi)-\xi\sigma].
\end{equation}
By solving it and substituting back to the action, we find that the contributions of $\psi$ cancel, leading us to a single massive scalar dof:  
\begin{equation}
    \mathcal{L}_S=-\frac{ 1}{2}\partial_{\mu}\sigma\partial^{\mu}\sigma-\frac{\xi^2}{16\beta}\sigma^2
\end{equation}
This is also intuitive -- the pure $R^2$ gravity has no dof in flat space-time \cite{Hell:2023mph}. Thus, 

\begin{equation}
    \textit{\textbf{Case 2:} $n>1$, or  $m>1$ and  $l=2$}    
\end{equation}
In this case, the non-minimal coupling does not contribute to the Lagrangian density. By varying the action with respect to the scalar $\phi$, we find the following constraint:
\begin{equation}
    8\beta\Delta\phi=-8\beta(3\ddot{\psi}-2\Delta\psi),
\end{equation}
which exactly cancels the contribution of $\psi$. Thus, we are left again with a single scalar mode, which is now massless, due to the absence of non-minimal coupling: 
\begin{equation}
    \mathcal{L}_S=-\frac{ 1}{2}\partial_{\mu}\sigma\partial^{\mu}\sigma-\lambda\sigma^s
\end{equation}

\begin{equation}
    \textit{\textbf{Case 3:} $n=1$,  $m=1$ and  $l>2$}    
\end{equation}
If instead we increase the power of the Ricci scalar, and keep $n=m=1$, then, at the leading order, we find:  
\begin{equation}\label{Ls0mode}
    \mathcal{L}_S=-\frac{ 1}{2}\partial_{\mu}\sigma\partial^{\mu}\sigma-\xi\sigma\left(\Delta\phi+3\ddot{\psi}-2\Delta\psi\right)
\end{equation}
Interestingly, now we can see that $\phi$ gives rise to the following constraint:
\begin{equation}
    \Delta\sigma=0\qquad\to\qquad \sigma=0. 
\end{equation}
Thus, at leading order, this case does not give rise to any scalar modes:
\begin{equation}
    \mathcal{L}_S=0. 
\end{equation}
One might be surprised at this result: we have started with the non-vanishing kinetic term for $\sigma$, and thus it would be unexpected to conclude that there are no dof. However, this mode is now coupled with the other two modes.  Among them, the first one is constrained, while the second coupling gives rise to the non-vanishing contribution to the kinetic matrix. 

An alternative way to see that there are no propagating scalar dof would be to decouple $\psi$ and $\sigma$. This can be done by the following substitution:
\begin{equation}
    \sigma=\omega+3\xi\psi, 
\end{equation}
after which (\ref{Ls0mode}) becomes:
\begin{equation}
    \mathcal{L}_S=-\frac{1}{2}\partial_{\mu}\omega\partial^{\mu}\omega-\frac{9}{2}\xi^2\dot{\psi}^2-\frac{3}{2}\xi^2\psi\Delta\psi-\xi\Delta\psi\omega+\Delta\phi(3\xi^2\psi-\xi\omega)
\end{equation}
Now the constraint is given by:
\begin{equation}
    \omega=3\xi\psi. 
\end{equation}
By substituting it back to the Lagrangian density, we find:
\begin{equation}
    \mathcal{L}_S=0, 
\end{equation}
which is consistent with the previous result on the absence of scalar modes in this case. 

\begin{equation}
    \textit{\textbf{Case 4:} $n>1$, or $m>1$ and  $l>2$}    
\end{equation}
Finally, if one of the powers $n$ or $m$ is larger than unity, and $l>2$, neither non-minimal coupling nor the Ricci scalar contributes to the scalar modes, and thus we remain only with the propagating scalar field $\sigma$. 

Therefore, we arrive at a striking and unexpected result: not only does the number of propagating modes change in the special cases, but if the non-minimal coupling enters the action linearly in the scalar field -- regardless of whether this field carries a kinetic term, or no, then, after taking into account all of the constraints, we find that the theory defined by the third case contains no dof at all at the leading order. Let us see if these results also hold in other cases. 

\section{Variable gravity and the power-law generalization}

As next set of models, we will consider theories of variable gravity, and their generalizations with higher powers of the Ricci scalar, which are described by the following action: 
\begin{equation}\label{actionVG}
     S_{vgg}=\int d^4x\sqrt{-g}\left[-\frac{ 1}{2}\nabla_{\mu}\sigma\nabla^{\mu}\sigma-\frac{\xi}{2}\sigma^nR^m- \lambda \sigma^s\right].
\end{equation}

\subsection{The Einstein frame}
Similarly to the pure case, this theory has no Einstein term. Let us first find its expression in the Einstein frame, in which it is going to emerge. For this, we introduce another scalar field:
\begin{equation}\label{vg2}
     S_{vgg}=\int d^4x\sqrt{-g}\left[-\frac{ 1}{2}\nabla_{\mu}\sigma\nabla^{\mu}\sigma- \lambda \sigma^s-\frac{\xi m (m-1)}{2}\sigma^n\left(\frac{1}{m-1}R\chi^{m-1}-\frac{1}{m}\chi^m\right)\right]
\end{equation}
By varying the action with respect to $\chi$, which is a constrained field, we find:
\begin{equation}
    \chi=R. 
\end{equation}
Clearly, $R=0$, $m=0$, and $m=1$ are singular points. Thus, in the following, we will assume that these do not hold, and return to them in the subsequent analysis. Let's next define a new metric:
\begin{equation}
    g_{\mu\nu}=-\frac{2 M^2}{\xi m}\sigma^{-n}\chi^{1-m} \Tilde{g}_{\mu\nu}
\end{equation}
and express the action (\ref{vg2}) in terms of it. At this point we should be careful: in order to preserve the signature of the metric, the overall coefficient should be positive: 
\begin{equation}
    -\frac{2 M^2}{\xi m}\sigma^{-n}\chi^{1-m}>0. 
\end{equation}
To avoid any confusion, we will thus define $\mu=-\xi$, and assume that the above condition is satisfied \footnote{Such condition should always hold, also for the previously considered models}. By further defining: 
\begin{equation}
    \Omega=M\ln\left(\frac{\mu m \sigma^n\chi^{m-1}}{M^2}\right),
\end{equation}
we find:
\begin{equation}
      S_{vgg}=\int d^4x\sqrt{-\Tilde{g}}\left\{M^2\Tilde{R}-\frac{3}{2}\Tilde{\nabla}_{\mu}\Omega\Tilde{\nabla}^{\mu}\Omega-    V_1(\Omega)\Tilde{\nabla}_{\mu}\sigma\Tilde{\nabla}^{\mu}\sigma+V_2(\sigma,\Omega)\right\},
\end{equation}
where 
\begin{equation}
    V_1(\Omega)=e^{-\frac{\Omega}{M}}\qquad \text{and}\qquad V_2(\sigma,\Omega)=-4\lambda e^{-\frac{2\Omega}{M}}\sigma^s-\frac{\mu(m-1)}{4}\left(\frac{M^2}{\mu m}\right)^{\frac{m}{m-1}}\sigma^{\frac{n}{1-m}}e^{\frac{(2m-1)\Omega}{(m-1)M}}
\end{equation}
Even though the two fields are coupled, the above action clearly describes two scalar fields and two tensor modes that arise from the Ricci scalar. This is clear if one linearizes the theory and notices that the two scalars decouple. 

\subsection{The singular points}
The previous transformation only holds for non-zero curvature, and if $m\neq0$ or $m\neq1$. If one takes $m=0$, one recovers the potential in powers of the scalar, which is already considered in the action, making this case thus trivial. 

\subsubsection{The flat background}
As the next step, let's consider the flat space-time background. Since there is no term linear in the Ricci scalar, at the leading order, there will be no tensor modes. This is also similar to the previous case, in which the tensor modes only appear with non-vanishing curvature. In addition, it is clear that the vector modes will be absent. The scalar modes will again be non-trivial. 

\begin{equation}
    \textit{\textbf{Case 1:} $n>1$, or  $m>1$ }    
\end{equation}

In this case, the scalar modes will always propagate, with no contribution from the non-minimal coupling. In particular, we find with $s=2$: 
\begin{equation}
    \mathcal{L}_{S,vgg}=\frac{  1}{2}\left(\dot{\sigma}^2+\sigma\Delta\sigma\right)-\lambda \sigma^2
\end{equation}

\begin{equation}
    \textit{\textbf{Case 2:}  m=1\qquad \text{and}\qquad n=1 }    
\end{equation}

 If, in contrast, this case holds, then we arrive at the following Lagrangian density for the scalar perturbations: 
\begin{equation}
    \mathcal{L}_{S,vgg}=\frac{  1}{2}\left(\dot{\sigma}^2+\sigma\Delta\sigma\right)-\lambda \sigma^s-\xi\sigma\left(\Delta\phi+3\ddot{\psi}-2\Delta\psi\right)
\end{equation}
We can see that $\phi$ is again not propagating. By varying the action with respect to it, we find:
\begin{equation}
    \Delta\sigma=0,
\end{equation}
which implies that:
\begin{equation}
    \sigma=0. 
\end{equation}
However, this means that the Lagrangian density for the scalar modes again vanishes, similarly to the previous section:
\begin{equation}
    \mathcal{L}_S=0. 
\end{equation}
Thus,the variable gravity with power-laws has no scalar modes at leading order in flat space-time, for $n=m=1$. Otherwise, it propagates a single scalar. 

\subsubsection{Homogeneous and isotropic background with $m=1$ }

Let us now instead consider the curved space-time, with $m=1$. For simplicity, we will first also set $n=1$, and then generalize to the case with general $n$.  In this case, we will assume the homogeneous and isotropic Universe, and a non-vanishing value of the scalar field, with conventions and perturbations according to the relations (\ref{FLRWbcg}) -- (\ref{FLRWsc}). 

First, we can find the background equations of motion by varying the action with respect to the lapse, the scalar field, and the scale factor. We then solve these equations for the coupling and the second derivatives of the scale factor and the scalar:
\begin{equation}
    \begin{split}
     \xi &= 
-\frac{     \dot{\sigma}^{2} a^{2}}{6 \dot{a} \left(a \dot{\sigma}+\sigma \dot{a}\right)}\\
\ddot{a} &= 
\frac{-5 a^{2} \dot{\sigma}^{2} \dot{a}^{2}-12 \sigma a \dot{\sigma} \dot{a}^{3}-8 \sigma^{2} \dot{a}^{4}}{\left(a \dot{\sigma}+2 \sigma \dot{a}\right)^{2} a}\\
\ddot{\sigma} &= 
-\frac{5 \left(a^{2} \dot{\sigma}^{2}+\frac{14 \sigma a  \dot{a} \dot{\sigma}}{5}+\frac{12 \sigma^{2} \dot{a}^{2}}{5}\right) \dot{\sigma} \dot{a}}{\left(a \dot{\sigma}+2 \sigma \dot{a}\right)^{2} a},
    \end{split}
\end{equation}
 assuming that they are always satisfied while we perform the analysis of the perturbations. 

To study the modes, we then expand the action up to the second order in the perturbations. Similarly to the previous cases, the scalar, vector, and tensor modes decouple, allowing us to analyze them separately. 

Let us first consider the scalar modes. In this case, we first perform integration by parts and find that the field $\phi$ is not propagating. By varying the action, we find its corresponding constraint. However, its form is similar to the flat space-time -- $\phi$ only appears to be linear in the action. We further simplify it by substituting:
\begin{equation}\label{redefvg}
     \psi=\psi_2+\frac{\dot{a}}{a\dot{\sigma}}\delta\sigma
\end{equation}
Then, we solve the constraint for $\delta\sigma$, and substitute this expression back to the action. 

At this point, $\phi$ drops out of the action, and we only remain with $\psi_2$. After performing several integrations by parts, we find:
\begin{equation}
    \mathcal{L}_{S,vgg}=\frac{     \dot{\sigma}^{2} a^{3} \sigma \left(a^{2} \dot{\psi}_2^{2}-\psi_2^{2} k^{2}\right)}{2 \dot{a} \left(a \dot{\sigma}+\sigma \dot{a}\right)}
\end{equation}
Therefore, in contrast to the flat space-time, the $n=1$ case with curved background propagates one scalar degree of freedom. 

In the case of tensor modes, after several integrations by parts, we find two degrees of freedom:
\begin{equation}
    \mathcal{L}_{T,vgg}=\frac{\sigma\dot{\sigma}^2 a^5   }{48\dot{a}(a\dot{\sigma}+\sigma\dot{a})}\left(\dot{h}_{ij}^T\dot{h}_{ij}^T+\frac{1}{a^2}h_{ij}^T\Delta h_{ij}^T\right)
\end{equation}

One can easily show that if $n\neq1$, the theory still propagates a single scalar and two tensor modes. In this case, for the scalar sector, one will encounter a similar procedure as in the $n=1$ case, with the scalar mode  re-defined as in (\ref{redefvg}), and with the corresponding Lagrangian density given by:
\begin{equation}
    \mathcal{L}_{S,vgg}=\frac{     \dot{\sigma}^{2} a^{3} \sigma \left(a^{2} \dot{\psi}_2^{2}-\psi_2^{2} k^{2}\right)}{2 \dot{a} \left(n a \dot{\sigma}+\sigma \dot{a}\right)}
\end{equation}

The Lagrangian density for the tensor modes in this case also generalizes to: 
\begin{equation}
    \mathcal{L}_{T,vgg}=\frac{\sigma\dot{\sigma}^2 a^5   }{48\dot{a}(na\dot{\sigma}+\sigma\dot{a})}\left(\dot{h}_{ij}^T\dot{h}_{ij}^T+\frac{1}{a^2}h_{ij}^T\Delta h_{ij}^T\right)
\end{equation}

Curiously, thus, the variable gravity case $m=1$ is similar to the pure scale-invariant gravity. In flat space-time, it has no dof, while in curved space-time, it propagates a scalar and two tensor modes. 

\section{Power-law curvature with the minimal coupling to the scalar}
In the previous cases, the scalar field was non-minimally coupled to the curvature. However, another possibility is that it is minimally coupled and described by the following action:  
\begin{equation}\label{minS}
     S_{minPS}=\int d^4x\sqrt{-g}\left[\frac{M^2}{2}\left(R-2\Lambda\right)-\frac{   1}{2}\nabla_{\mu}\sigma\nabla^{\mu}\sigma- \lambda \sigma^s+\beta R^l\right]. 
\end{equation}
If $\beta=0$, the above case simply reduces to Einstein gravity, coupled to the scalar field, which has three degrees of freedom: two tensor modes and a scalar one. Let us see how this changes if $M=0$, or if both $M$ and $\beta$ are non-vanishing. 

\subsection{The general case}

Let us first consider the case when both $M\neq 0$, and $\beta\neq 0$. By following the procedure analogous to the case of the power-law gravity, we can express (\ref{minS}) in the Einstein frame: 
\begin{equation}
     \begin{split}
           S_{minPS}&=\int d^4x\sqrt{-\tilde{g}}\left\{M^2\tilde{R}-\frac{3}{2}\tilde{\nabla}_{\mu}\Omega\tilde{\nabla}^{\mu}\Omega-    e^{-\Omega/M}\tilde{\nabla}_{\mu}\sigma\tilde{\nabla}^{\mu}\sigma\right.\\&\left. +4e^{-2\Omega/M}\left[-M^2\Lambda-\lambda\sigma^s+\beta^{\frac{1}{1-l}}(1-l)\left[\frac{M^2}{2}\left(e^{\frac{\Omega}{M}}-1\right)\right]^{\frac{1}{l-1}}\right]\right\}, 
     \end{split}
\end{equation}
where $\Omega$ was defined as in (\ref{OmegaDef}), and
\begin{equation}
    F=\frac{M^2}{2}+\beta l \chi^{l-1}. 
\end{equation}
The above theory clearly propagates two scalar modes. One is already present in the theory, while the other arises due to the higher powers in the curvature. 

Thus, in general, we have found four degrees of freedom: in addition to the scalars, we have two tensor modes that are contained in the Einstein term. However, as we have previously seen, the structure of the dof can change if one instead considers the Jordan frame, with a flat background. In this case, one will always have two tensor modes, due to the appearance of the term linear in the Ricci scalar. In contrast, the number of scalars can change. By perturbing around the flat space-time, and decomposing the metric as in (\ref{decompositionFlat}), we find the following Lagrangian density for the scalar modes: 
\begin{equation}
    \begin{split}
        \mathcal{L}_S=M^2\left[2\phi\Delta\psi-3\dot{\psi}^2-\psi\Delta\psi\right]-\frac{   1}{2}\sigma_{,\mu}\sigma^{,\mu}-\lambda \sigma^s+2^l\beta\left(\Delta\phi+3\Ddot{\psi}-2\Delta\psi\right)^l
    \end{split}
\end{equation}
Let us study it depending on different cases. 

\begin{equation}
    \textit{\textbf{Case 1:}   $l=2$}    
\end{equation}
In this case, the $R^2$ term contributes to the scalar modes. Similarly to the previous cases, the scalar field $\phi$ is constrained and satisfies the following equation:
\begin{equation}
    8\beta\Delta\phi=-8\beta(3\ddot{\psi}-2\Delta\psi).
\end{equation}
By solving it and substituting back into the action, we find two decoupled scalar modes with the following Lagrangian density:
\begin{equation}
    \mathcal{L}_S=-\frac{   1}{2}\sigma_{,\mu}\sigma^{,\mu}-\lambda \sigma^s+3M^2\left(\dot{\psi}^2-\phi\Delta\psi-\frac{M^2}{4\beta}\right)
\end{equation}

\begin{equation}
    \textit{\textbf{Case 2:}   $l>2$}    
\end{equation}
If, on the other hand, $l$ takes values larger than 2, then the higher powers of the Ricci scalar do not contribute to the scalar modes. This means that the constraint for $\phi$ becomes:
\begin{equation}
    \Delta\psi=0\qquad\to\qquad \psi=0,
\end{equation}
and thus we remain with only one scalar mode: 
\begin{equation}
    \mathcal{L}_S=-\frac{   1}{2}\sigma_{,\mu}\sigma^{,\mu}-\lambda \sigma^s
\end{equation}
\subsection{The pure gravity case}
Let us now set $M=0$. Then, we are in the pure-gravity scenario, where the term linear in the Ricci scalar is absent. To go to the Einstein frame, we follow a procedure similar to \cite{Alvarez-Gaume:2015rwa, Hell:2025wha}, and introduce another scalar field: 
\begin{equation}
    S=\int d^4x\sqrt{-g}\left(-\frac{   1}{2}\nabla_{\mu}\sigma\nabla^{\mu}\sigma-\lambda\sigma^s+\beta l (l-1) \left[\frac{1}{l-1}R\chi^{l-1}-\frac{1}{l}\chi^l\right]\right)
\end{equation}
By varying the action with respect to it, we find:
\begin{equation}
    \chi=R. 
\end{equation}
By expressing the metric as:
\begin{equation}
    g_{\mu\nu}=\frac{M^2}{F}\tilde{g}_{\mu\nu},\qquad \text{where}\qquad F=\frac{\beta}{l-1}\chi^{l-1}
\end{equation}
and further defining $\Omega$, as in (\ref{OmegaDef}), we arrive at the following action:
\begin{equation}
     \begin{split}
           S_{minPS}&=\int d^4x\sqrt{-\tilde{g}}\left\{M^2\tilde{R}-\frac{3}{2}\tilde{\nabla}_{\mu}\Omega\tilde{\nabla}^{\mu}\Omega-     e^{-\Omega/M}\tilde{\nabla}_{\mu}\sigma\tilde{\nabla}^{\mu}\sigma\right.\\&\left. +4e^{-2\Omega/M}\left[-\lambda\sigma^s+\beta^{\frac{1}{1-l}}(l-1)\left[\frac{M^2(l-1)}{2}e^{\frac{\Omega}{M}}\right]^{\frac{l}{l-1}}\right]\right\}, 
     \end{split}
\end{equation}
We can see that for $l=2$, the last term becomes a constant. Thus, we arrive at an action describing two tensor fields and two scalar modes. 

Let us now go back to the Jordan frame to explore the $R=0$ case. At leading order, there are clearly no tensor modes. However, there will be only a single scalar mode, arising from $\sigma$ as well, and no contribution from the Ricci scalar. In order to see this, let us consider the Lagrangian for the scalar modes: 
\begin{equation}
    \begin{split}
        \mathcal{L}_S=-\frac{1    }{2}\sigma_{,\mu}\sigma^{,\mu}-\lambda \sigma^s+2^l\beta\left(\Delta\phi+3\Ddot{\psi}-2\Delta\psi\right)^l
    \end{split}
\end{equation}
We can see that at leading order, $\phi$ will always be constrained, and satisfy:
\begin{equation}
    \Delta\phi=-(3\Ddot{\psi}-2\Delta\psi)
\end{equation}
However, if we substitute this back into the action, we find that the contribution from the higher powers of the Ricci scalar vanishes, leaving us with a single scalar dof ($s=2$): 
\begin{equation}
    \begin{split}
        \mathcal{L}_S=-\frac{1    }{2}\sigma_{,\mu}\sigma^{,\mu}-\lambda \sigma^2
    \end{split}
\end{equation}
\section{Discussion and Summary}
Theories of gravity that involve powers of the Ricci scalar and scalar fields, together with their non-minimal couplings, have long been central to several areas of physics, including cosmology, modified theories of gravity, supergravity, and string theory. Such extensions are expected to emerge naturally from fundamental physics, either through vacuum fluctuations or compactifications of extra dimensions. At the same time, they are also of strong observational relevance and have gained significant attention in light of recent DESI, ACT, and SPT data releases.

A particularly important aspect of these theories lies in their degrees of freedom (dof). These building blocks encode crucial information about the perturbative regime, the connection to observations, and the possible breakdown of perturbation theory due to strong coupling effects. \textcolor{Black}{In this work, we have addressed these questions through a systematic analysis. While the Einstein frame provides a relatively general framework for studying the modes, it does not capture all possible cases. In particular, we have identified singular points for which the Jordan frame analysis becomes more appropriate. As a result, a complete understanding of the theory requires working in both frames, which can lead to surprising conclusions regarding the number of propagating modes, as we will summarize in the following tables.
}

In the first case, we have performed the analysis for a general theory, which involves different powers of the Ricci scalar, together with its corresponding linear term (GR), and a non-minimal coupling to the scalar field. Our results can be summarized in Table 1.
\begin{table}[ht]
  \centering
  \ra{1.3} 
  \begin{tabular}{@{} l @{\hspace{1cm}} l @{}}
    \toprule
    {The case} & {Dof in $S_{PGS}$}  \\
    \midrule
    General case    & 2 tensor modes, and 2 scalar modes  \\
    $\eta_{JF\mu\nu}:$ $m=1,$ $n=1$ \& $l=2$ & 2 tensor modes, and 2 scalar modes \\
     $\eta_{JF\mu\nu}:$ $m>1,$ or $n>1$ \& $l=2$ & 2 tensor modes, and 2 scalar modes  \\
     $\eta_{JF\mu\nu}:$ $m=1,$ $n=1$ \& $l>2$ & 2 tensor modes, and 1 scalar modes  \\
     $\eta_{JF\mu\nu}:$ $m>1,$ or $n>1$ \& $l>2$ & 2 tensor modes, and 1 scalar modes  \\
    \bottomrule
  \end{tabular}
  \caption{The summary of the results for the general action (\ref{action1}). $\eta_{JF\mu\nu}$ stands for the evaluation of the theory for a Minkowski space-time in the Jordan frame. }
  \label{tab:comparison}
\end{table}
First, we note that in the general case, in which $R\neq$0, the theory describes two tensor modes and two scalar modes. While we have shown this in the Einstein frame, the same also holds in the Jordan frame (JF). However, the results start to diverge when we consider the Jordan frame with a Minkowski background, denoting it in the above table as $\eta_{JF\mu\nu}$. This point is singular, as it corresponds to the vanishing scalar $\chi$. The appearance of the second scalar $\psi$ in this case is clearly due to the quadratic powers in the Ricci scalar, as it is absent for $l\neq 2$, thus leading to a different number of dof.

As a second case, we have removed the terms non-linear in the Ricci scalar alone, with results summarized in Table 2. 
\begin{table}[ht]
  \centering
  \ra{1.3} 
  \begin{tabular}{@{} l @{\hspace{1cm}} l @{}}
    \toprule
    {The case} & {Dof in $S_{Enm}$}  \\
    \midrule
    General case    & 2 tensor modes, and 2 scalar modes  \\
    $\eta_{JF\mu\nu}:$ $m>1,$ or $n>1$  & 2 tensor modes, and 1 scalar mode \\
     $\eta_{JF\mu\nu}:$ $m=1,$ and $n=1$  & 2 tensor modes, and 1 scalar mode \\
     $g^{HI}_{\mu\nu}:$ $m=1,$ $n$ arbitrary & 2 tensor modes, and 1 scalar mode  \\
    \bottomrule
  \end{tabular}
  \caption{The summary of the results for the case of Einstein gravity with the non-minimally coupled scalar, described by the action (\ref{Enm}). $g^{HI}_{\mu\nu}$ denotes the homogeneous and isotropic background, while $\eta_{JF\mu\nu}$ corresponds to the flat space-time case, which we have evaluated in the Jordan frame (and for which the corresponding Einstein frame was singular)}
  \label{tab:comparison}
\end{table}
Similarly to the previous theory, the general case again assumes that $R\neq$0, with now in addition $m\neq1$. Otherwise, one encounters singular points, which we have analyzed in the Jordan frame, and another Einstein frame for $m=1$ to additionally confirm the behavior. Notably, in these points, the number of propagating scalar modes changes when compared to the general case at the leading order in perturbations.

If, instead, one chooses to drop the linear term, as is the case in pure gravity with non-minimal couplings, whose results are summarized in Table 3, or variable gravity and its generalizations, which are summarized in Table 4, one is led to unexpected results. Dof stinkingly changes depending on the parameters, admitting even the case when theories have no propagating modes. 

\begin{table}[ht]
  \centering
  \ra{1.3} 
  \begin{tabular}{@{} l @{\hspace{1cm}} l @{}}
    \toprule
    {The case} & {Dof in $S_{Hnm}$}  \\
    \midrule
    General case    & 2 tensor modes, and 2 scalar modes  \\
    $\eta_{JF\mu\nu}:$ $m=1,$ $n=1$ \& $l=2$  &  1 scalar mode \\
     $\eta_{JF\mu\nu}:$ $m=1,$ $n=1$ \& $l>2$  &  no modes \\
     $g^{HI}_{\mu\nu}:$ $m>1,$  or $n>0$, \& $l>2$ & 1 scalar mode  \\
    \bottomrule
  \end{tabular}
  \caption{The summary of the results for the case of higher-power curvature with the non-minimally coupled scalar, described by the action (\ref{Pnm}), and following the conventions of the previous tables. The general case assumes that $R\neq$0, and holds in both Jordan and Einstein frames, while the special cases assume $R=0$, and evaluate the theory in the Jordan frame. }
  \label{tab:comparison}
\end{table}

\begin{table}[ht]
  \centering
  \ra{1.3} 
  \begin{tabular}{@{} l @{\hspace{1cm}} l @{}}
    \toprule
    {The case} & {Dof in $S_{vgg}$}  \\
    \midrule
    General case    & 2 tensor modes, and 2 scalar modes  \\
    $\eta_{JF\mu\nu}:$ $m>1,$ or  $n>1$   &  1 scalar mode \\
     $\eta_{JF\mu\nu}:$ $m=1,$ \& $n=1$   &  no modes \\
     $g^{HI}_{\mu\nu}:$ $m=1,$  \& arbitrary n & 1 scalar mode and 2 tensor modes  \\
    \bottomrule
  \end{tabular}
  \caption{The summary of the results for the case of Variable gravity and the power-law generalization, described by the action (\ref{actionVG}), and following the conventions of the previous tables. The general case assumes that $R\neq$0, and holds in both Jordan and Einstein frames, while the special cases assume $R=0$, or $m=1$. } 
  \label{tab:comparison}
\end{table}

This behavior is especially surprising as one starts with a propagating scalar field. However, due to the non-minimal coupling with the Ricci scalar, in the $m=1$ and $n=1$ case, this contribution is irrelevant. The main reason for this is the constrained $h_{00}$ component -- the gravitational potential $\phi$, which never appears with time-derivatives, unlike the other gravitational scalar $\psi$. As a result, its constraint sets the external scalar $\sigma$ to zero, regardless of the presence of the kinetic term, which in turn influences also the propagation of $\psi$, ultimately vanishing. 

For completeness, in addition to the non-minimal couplings, we have also considered a theory in which the scalar is minimally coupled, with the results summarized in Table 5. Similarly to the previous cases, this theory also exhibits a jump in the dof, depending on different cases. 
\begin{table}[ht]
  \centering
  \ra{1.3} 
  \begin{tabular}{@{} l @{\hspace{1cm}} l @{}}
    \toprule
    {The case} & {Dof in $S_{minPS}$}  \\
    \midrule
    General case    & 2 tensor modes, and 2 scalar modes  \\
    $\eta^G_{JF\mu\nu}:$ $l=2$   &  2 tensor modes, and 2 scalar modes \\
     $\eta^G_{JF\mu\nu}:$ $l>2$   &  2 tensor modes, and 1 scalar mode \\ 
     PG General case    & 2 tensor modes, and 2 scalar modes  \\  
     $\eta^{PG}_{JF\mu\nu}:$ $l>2$   &  1 scalar mode  \\  
    \bottomrule
  \end{tabular}
  \caption{The summary of the results for the case of power-law curvature with the minimal coupling to the scalar, given in (\ref{minS}), and following the conventions of the previous tables. The general case assumes that $R\neq$0, and holds in both Jordan and Einstein frames, while the special case $\eta^G_{JF\mu\nu}$ assumes $R=0$. Both of these consider $M\neq0$ and $\beta\neq0$. PG general stands for the pure gravity, minimally coupled to the scalar field.  } 
  \label{tab:comparison}
\end{table}

\textcolor{Black}{
Therefore, among the above cases, we can notice the following trend: the theories with non-minimal coupling between the scalar field and the curvature in curved space-time will essentially lead to two scalar modes and two tensor modes, which can be nicely seen in the Einstein frame. This, however, is not always the case, as there are exceptions for which the Jordan frame is most suitable for performing the analysis. Among these special cases, there are two that show particularly intriguing behaviour in flat space-time -- the case of variable gravity with $m=1$ and $n=1$, and the pure gravity with non-minimally coupled curvature, which in addition has $l>2$. The most surprising property of such theories is that, to leading order, they describe no propagating modes. This is similar to the pure $R^2$ gravity, but nevertheless surprising, because in both cases we have started with a propagating scalar field, which then disappears due to the constraint arising from the metric perturbation. In addition to this, Jordan frame analysis is special overall -- it captures all the effects found in the Einstein frame, as well as all singular points, which, as we can see, have a different number of dof. }

\textcolor{Black}{It should be also noted that among} the possible gauge fields with non-minimal coupling to gravity, the scalar field appears to be special. For example, the non-minimal couplings between a vector field and Ricci scalar can give rise to the extra dof, with vanishing speed of propagation, which indicates a breakdown of perturbation theory \cite{DeFelice:2025ykh}. If, in addition, coupled to the Ricci tensor, one of the branches of the theory can lead to the strong coupling of tensor modes, unless defined in a disformal frame \cite{Hell:2024xbv}. The 3-form with non-minimal coupling does not give rise to new dof. However, similarly to the vector field, it has multiple branches, emerging for an anisotropic universe, with one containing modes whose speed of propagation vanishes \cite{DeFelice:2025khe}. This thus makes the scalar field special --\textcolor{Black}{in the cases that we were studying, the modes were always propagating with unit speed of propagation}. 

\textcolor{Black}{At the same time, the different number of dof indicates that the general theories which propagate two scalar and two tensor modes above could have strong couplings, in the limit when the curvature vanishes. This is similar to the case of $R^2$ gravity, or massive gauge theories in general \cite{Hell:2023mph, Hell:2025uoc}. Thus, one might expect that the modes that are appear as soon as one moves away from the special cases like flat space-time, or $m=1$ case in the variable gravity, could become strongly coupled at a scale analogous to the Vainshtein radius in massive gravity, and decouple from the remaining dof beyond it. Nevertheless, at this point, we will take this possibility simply as an indication, and perform the strong-coupling analysis in the future work.  }

\textcolor{Black}{ In addition to considering cases where the scalar $\sigma$ as its associated kinetic term, there is also a special case, where it is constrained. As pointed out in \cite{Hell:2025lgn}, for a special case, such theory has an interesting structure as well, being able to admit healthy fields and, at the same time, equation of state that drops below $-1$, among other curious solutions. While in this work we have considered only the cases in which the scalar field is propagating, investigating the cases with constrained scalars will also pose an interesting future study.  }

Overall, we can conclude that power-law gravity with non-minimally coupled scalars has an intriguing structure. While the question -- \textit{What is the number of degrees of freedom?} -- could seem particularly simple, this theory can nevertheless leave us with surprises, and signal awareness to the singular cases, appearing when studying modified theories of gravity in simpler formulations.

\begin{center}
\textbf{\textsc{Acknowledgments}}
\end{center}
\textit{A. H. would like to thank the Max Planck Institute for Physics (MPP)
in Garching for its hospitality during
her visit, when part of this work was carried out. The work of A. H. was supported in part by JSPS KAKENHI No.~24K00624, and 
by the World Premier International Research Center Initiative (WPI), MEXT, Japan. The work of D.L. is supported by the Origins Excellence Cluster and by the German-Israel-Project (DIP) on Holography and the Swampland.}

\bibliographystyle{utphys}
\bibliography{paper}

\end{document}